\documentclass{egpublarxiv}
\usepackage{eg2018}

\ConferenceSubmission
\electronicVersion
\PrintedOrElectronic

\usepackage[pdftex]{graphicx} \pdfcompresslevel=9

\def\mysection#1#2{\section{#1}\label{sec:#2}}
\def\mysubsection#1#2{\subsection{#1}\label{sec:#2}}

\usepackage{t1enc,dfadobe}
\usepackage{soul}
\usepackage{amsmath}
\usepackage{amssymb}
\usepackage{microtype}
\usepackage{wrapfig}
\usepackage{todonotes}

\def\figurePath{images/}
\def\myfigure#1#2{\begin{figure}[htb]\centering\includegraphics*[width = \linewidth]{\figurePath#1}\caption{#2}\label{fig:#1}\end{figure}}

\def\mycfigure#1#2{\begin{figure*}[t]\centering\includegraphics*[clip, width = \linewidth]{\figurePath#1}\caption{#2}\label{fig:#1}\end{figure*}}

\newcommand{\eg}{e.\,g.\ }
\newcommand{\ie}{i.\,e.\ }
\newcommand{\etal}{et~al.\ }

\newcommand{\refSec}[1]{Sec.~\ref{sec:#1}}
\newcommand{\refFig}[1]{Fig.~\ref{fig:#1}}
\newcommand{\refEq}[1]{Eq.~\ref{eq:#1}}

\newcommand{\change}[1]{#1}

\soulregister\ref7
\soulregister\cite7
\soulregister\refFig7
\soulregister\cite7
\soulregister\ref7
\soulregister\pageref7
\soulregister\shortcite7
\soulregister\eg0
\soulregister\ie0
\soulregister\etal0

\DeclareGraphicsExtensions{.png,.jpg,.pdf,.ai,.psd}
\newcommand{\paper}

\title{Single-image Tomography: 3D Volumes from 2D Cranial X-Rays}

\author[Henzler et al.]
	{\parbox{\textwidth}{
    \centering 
    Phlipp Henzler$^{1}$ \qquad
    Volker Rasche$^{1}$ \qquad 
    Timo Ropinski$^{1}$ \qquad     
    Tobias Ritschel$^{2}$}
    \\
	{
    \parbox{\textwidth}{\centering 
    	$^1$Ulm University, Germany
        \qquad
        $^2$University College London, UK}
		}
	}

    \teaser{
\includegraphics[width=\linewidth]{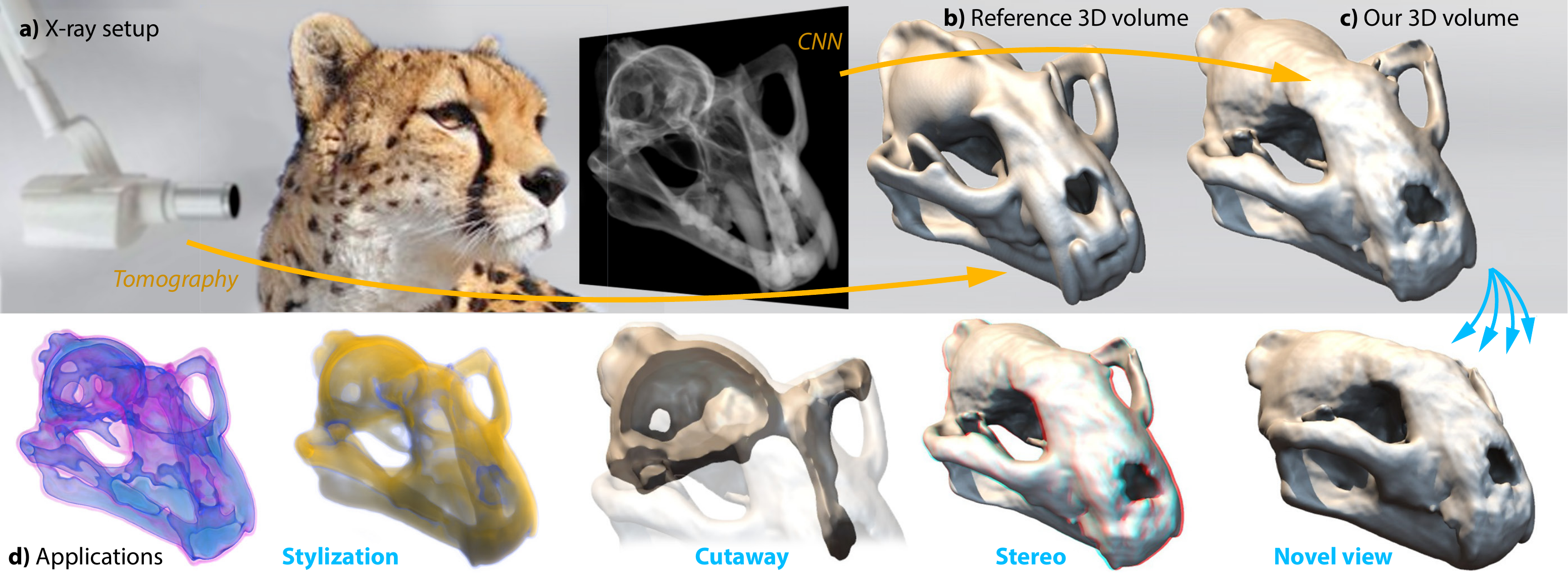}
\centering
\caption{
X-ray equipment \emph{(a)} has widespread availability enabling the creation of 2D x-ray imagery (here of a Cheetah) at low cost and with high speed. 
Regrettably, 2D x-rays do not reveal the full 3D structure \emph{(b)}, here visualized as an iso-surface, available to equipment such as a CT scanner.
Our deep learning-based result recovers the full 3D density volume \emph{(c)} from a single x-ray, allowing for \protect\change{prototypical} applications \emph{(d)} such as cut-away to reveal the \protect\change{cranial} interior,
re-rendering under different viewpoints, or
stereoscopic rendering of legacy x-ray material \protect\change{(here red-cyan anaglyph) or stylized-re rendering in a new modality, here a colorful transfer function}.}
\label{fig:Teaser}
}

\begin{document}

\maketitle 

\begin{abstract}
As many different 3D volumes could produce the same 2D x-ray image, inverting this process is challenging.
We show that recent deep learning-based convolutional neural networks can solve this task.
As the main challenge in learning is the sheer amount of data created when extending the 2D image into a 3D volume, we suggest firstly to learn a coarse, fixed-resolution volume which is then fused in a second step with the input x-ray into a high-resolution volume.
To train and validate our approach we introduce a new dataset that comprises of close to half a million computer-simulated 2D x-ray images of 3D volumes scanned from 175 mammalian species.
\change{Future} applications of our approach include stereoscopic rendering of legacy x-ray images, re-rendering of x-rays including changes of illumination, view pose or geometry.
Our evaluation includes comparison to previous tomography work, previous learning methods using our data, a user study and application to a set of real x-rays.
\end{abstract}

\mysection{Introduction}{Introduction}

Producing 2D images of a 3D world is inherently a lossy process, \ie the entire geometric richness of 3D gets projected onto a single flat 2D image. Consequently, any attempt to undo this operation is a daunting task. X-ray, or any other volumetric imaging technique is not different in this respect from photography of opaque surfaces. However, while x-ray enables us to ``see inside'' a solid object the spatial structure is only apparent to an expert with previous experience, typically in analyzing medical imaging data. On the one hand, inverting x-ray imagery might be more difficult than inverting opaque 2D images, as multiple transparent observations mix into a single pixel, following the intricate laws of non-linear volumetric radiation transport. On the other hand, there is also hope that semi-transparent imaging fairs better compared to solid-surface imaging as occlusion is not binary and additional surfaces remain accessible.

In this work, we apply deep learning in the form of convolutional neural networks (CNNs) to the challenge of inverting x-ray imagery. While CNNs have had success in generating depth from opaque observations~\cite{eigen2014depth} and inferring full 3D volumes~\cite{wu20153d,rezende2016unsupervised,Firman2016Structured}, we are not aware of any attempts to invert single x-ray images, or  other transparent modalities. 

\myfigure{Educational}{Differences between previous work and our approach.}
\refFig{Educational} conceptualizes the difference to previous work in flatland: The first CNNs (brain icon) consumed a 2D image to output a 2D height field representation of the 3D world as shown in the first column.
Later work has considered binary voxelizations shown in the middle column. Ours, in the last column, addresses transparent surfaces and produces results with continuous density, both essential for x-rays as found in \change{medical} applications or the natural sciences.

Performing this mapping is an important challenge as the quantity of x-ray images for which the real 3D volume is unknown, lost or inaccessible is likely substantial. An example would be a large repository of x-ray imagery acquired before 3D imaging such as CT scanning was invented. If this legacy material can be made to ``become 3D'' again, many interesting computer graphics applications become possible. Our results indicate
that 3D volumes can be inferred from 2D x-ray imagery for a certain quality level for a specific class of inputs. In this paper, we focus on x-rays of mammalian anatomical structures, typically crania (cf.\ \refFig{Teaser}, a), and demonstrate previously impossible computer graphics applications, such as volumetric cut-aways, novel view synthesis, stylization or stereoscopic rendering (\refFig{Teaser}, d).
We believe application to other content, such as x-ray for security or \change{medical applications might be} possible in future work, given training data is provided. 

Beyond the area of computer graphics, we also see the presented approach as a first important step towards applications in the life sciences. We envision \change{our work as a first step towards} a diagnostic tool in conditions where either no CT equipment, or the education to interpret x-ray imagery is available, such as for mobile x-ray devices, lay users, or \change{medical} diagnostics in developing countries. While there can be more than a hundred CT device units per one million inhabitants in industrial countries, the number is below one per million \eg in Africa~\cite{who2011survey}.
\change{
While the system described here is a first step towards this goal, evaluation in the presence of anomalies (pathologies), for different (non-cranial) classes and user studies with medical experts remain future work.}

To this end, our main contribution is two-fold: firstly, a new dataset that contains a large number of pairs of simulated 2D x-ray images and their corresponding 3D volumes forming a sampling of the mapping we wish to invert. Secondly, an investigation of an approach involving a CNN architecture to learn this inverse mapping so that it can be applied to legacy 2D x-ray images, as well as a fusion step to make it scale to large resolutions to overcome limitations of learning. Our evaluation of the proposed network architecture includes quantitative measurements with respect to the given data set and comparison to learning and non-learning-based baselines, qualitative evaluations on real x-ray imagery where no ground truth is available and a user study.


\mysection{Previous Work}{PreviousWork}
The problem at hand falls into the class of \emph{inverse rendering}. Instead of generating a 2D rendering from a 3D volume, as done in classical direct volume rendering~\cite{drebin1988volume}, we aim at an opposite challenge: how to obtain a 3D volume from a 2D x-ray image. While typically, such inverse problems are cast as an optimization procedure, such as based on deconvolution~\cite{ramamoorthi2001signal}, we suggest an approach purely based on learning from synthetic example data.

Typically, a volume is reconstructed from multiple views by means of computed \emph{tomography}~\cite{hounsfield1973computerized}. While the quality of tomographic reconstructions has increased by making use of principles such as maximum likelihood~\cite{shepp1982maximum} or sparsity~\cite{lustig2007sparse}, they typically require a large number of input images. Prior work on single-image tomography has used \change{statistical shape models \cite{lamecker2006atlas,novosad2004three}  or structural priors \cite{serradell2011simultaneous} of known anatomical structures}.
While these approaches deliver precise results, they are only applicable for specific problems, namely those where the content of the x-ray image is known. Our dataset however, contains many different anatomical structures from a multitude of different species in different poses.

In the computer graphics community, tomographic reconstruction methods for specific volumetric phenomena, such as flames~\cite{ihrke2004image} or planetary nebulae~\cite{magnor2004constrained}, have been proposed. Notably, the method of Magnor ~\etal~\cite{magnor2004constrained} works on a single image, but reconstructs only a very general shape, with a-priori known radial symmetry and an emission-only model. Instead, we employ deep learning~\cite{lecun2015deep}, in particular CNNs~\cite{jia2014caffe} to solve the task at hand. The presented approach has been inspired by previous work of Eigen~\etal which is generating a 2D depth map from a 2D image by using a two-staged CNN~\cite{eigen2014depth}. The next logical step following the reconstruction of 2D depth maps is to infer complete 3D volumes~\cite{wu20153d,rezende2016unsupervised}. Wu~\etal \change{, Choy\etal \cite{choy20163d} and Girdhar\etal\cite{girdhar2016learning}} infer a small binary 3D volume from 2D photos. 
\change{Tatarchenko \etal\cite{tatarchenko2017octree} and Fan\etal\cite{fan2016point} address the spatial resolution issue by regressing octrees, respectively point clouds instead of regular grids.}
In comparison to our task, this is harder (as parts are occluded), but at the same time easier (as one pixel still has a unique depth). 
\change{Wu and colleagues \cite{wu2016learning} as well as \cite{wang2017shape}
have devised adversarial designs to encode 3D volumes of shape classes allowing for single-image-to-3d applications.} 
While Qi~\etal have more recently improved upon the main task to make use of this approach (object recognition)~\cite{qi2016volumetric}, we are not aware of any generalizations, neither for semi-transparent surfaces nor for non-binary densities as it is required for instance for x-ray illustrated in \refFig{Educational}.

A typical challenge for deep learning is to find a suitably large number of training examples. We address this by using synthetic imagery in combination with real world volumetric CT scans. Since synthetic images can be used to enable tasks such as object detection at a similar quality as real data could~\cite{pepik2015holding}, they have been used to apply deep learning to classical problems such as optical flow~\cite{dosovitskiy2015flownet}, intrinsic images~\cite{narihira2015direct} or light and material estimation \cite{rematas2016deep}. In our case, the volumetric data has been collected from a large-scale database of mammalian CT scans which contains a large variety of skulls ranging from mole rats, over polar bears to chimpanzees and walruses. \refFig{Samples} shows a small subset of the synthetic x-rays and the corresponding volumes contained in our data set. While anatomy can be considered an important use case of x-rays, the transfer to other domains, such as airport security x-ray imaging, is left open for future work.

\mycfigure{Samples}{Samples of our dataset comprising of synthetic 2D x-ray images \emph{(top row)} and 3D volumes \emph{(bottom row)}.
The second row shows images rendered from the original viewpoint while the mapping is between view-dependent 2D images and view-independent 3D volumes.
}

Medical visualization has adapted deep learning for tasks that have similarity with ours. W\"urfl~\etal\cite{wurfl2016deep} have mapped tomographic reconstruction to the back-propagation~\cite{rumelhart1988learning} of a CNN. Their input is an x-ray sinogram, that captures a volume from many views, while our input is a single view only. While exploiting similarity of back-propagation in learning and tomography is an interesting observation, it does not make use of the main feature of deep learning: finding useful internal representations to invert a mapping. Hammerdink~\etal~\cite{hammernik2017deep} consider the special case of limited-angle tomography and used deep learning to remedy aliasing problems in reconstruction from very few images. Bahrami~\etal~\cite{bahrami2016convolutional} learn the mapping from 7T~CT to 3T~CT, whereby input and output is 3D, whereas we map from 2D to 3D. However, the high-level objective is similar: make most of the available data by means of deep learning reducing capture cost and effort. Our approach is the limit case: reconstruction from a single image.

\mysection{Dataset}{Dataset}
We chose to explain our dataset previous to the introduction of the method that makes use of it, as we hope it to be general enough to also serve other purposes in the future. 

Our dataset contains samples of the mapping from 2D x-ray images to 3D volume data. This represents both the forward and inverse mapping. We name a pair of 2D x-ray images and 3D volume a \emph{sample}. Examples of samples are shown in \refFig{Samples}. Overall we have produced 466,200 such samples, whereby computer-generated (simulated) x-ray imaging is used to achieve such a high number of x-ray images based on real-world CT volumes. The 3D volumes come from a repository of CT scans of Mammalia~\cite{UTCT}. Out of these 53,280 samples are withheld for validation. The validation set exclusively contains 20 species which are never observed at training time. Next, we will describe how we have obtained the data (\refSec{3DData}) and detail our approach for synthetic x-ray generation (\refSec{2DData}).

\mycfigure{Architecture}{
Input to our architecture is a 2D x-ray gray image \emph{(left)}.
The network converts this image into an internal representation with decreased spatial resolution (here seen as a block's height) and increasing depth (depicted as a block's width).
Each type of block (encoded as colors) is defined as a combination of other blocks.
Solid lines are learned, dotted lines are non-learned.
For details, please see the text.
}

\mysubsection{3D Data: Density Volumes}{3DData}
All data is \change{acquired from} the UTCT database \cite{UTCT} category ``Mammalia''. We have downloaded 175 slice videos in a resolution of approx.\ 500 pixels horizontally, acquired at varying quality, all subject to video compression, in 8 bit and with little calibration information, \ie in a rather unconstrained setting. We assume pixel values in these videos to be in units of linear density, \ie not to be subject to a gamma curve. All slice videos were re-sampled into volumes with a resolution of $128 \times 128 \times 128$ using a Gaussian reconstruction filter.
\change{Note that, as this is a very pragmatic acquisition process, producing training at a fidelity data that is below what a clean 3D scan will deliver, it remains future work to see what a network trained on clean data can do.} 
Finally, the 3D volume is re-sampled from different views (according to the x-ray image view), resulting in one unique volume per sample. 

Some of those species only have minor differences, such as female and male and others are completely different.
Similarly, species such as the Walrus or Koala in the validation can be very different from any species in the test data.


\mysubsection{2D Data: X-ray Images}{2DData}
In order to simulate real-world x-ray imaging, our image formation follows the Beer-Lambert absorption-only model. Intensity attenuation for each ray is simulated depending on the medium's density but never reflected. This is typical for x-rays~\cite{driggers2003encyclopedia}, as most relevant (organic) materials have an index of refraction very close to $1$ at x-ray wavelengths. Formally, the fraction $\alpha\in\mathbb R$ of x-radiation arriving at the sensor (transparency) after traveling a volume with extinction coefficient $\chi=\kappa+\sigma$ (the sum of absorption $\kappa$ and out-scattering $\sigma$) with spatially-varying density $\mu(t)\in\mathbb R^+$ along a ray parametrized with variable $t$, is
\begin{equation}
\alpha=
\exp
\left(
-
\chi\int\mu(t) \mathrm d t
\right)
\approx
\exp
\left(
-
\chi
\sum_{i=0}^{n_\mathrm s}
\mu(t_i)
\right).
\label{eq:Transparency}
\end{equation}

We have manually chosen $\chi$ globally to be $10$, which produces x-ray images with plausible contrast. Note that traditionally as well as in this work, an x-ray is represented not by means of transparency, but inverted by means of opacity which is defined as $1-\alpha$.

To generate the synthetic x-ray data, \ie to obtain transparency values for each pixel, a ray is marched front-to-back in $n_\mathrm s=128$ steps, solving $\alpha$ for a known $\chi$ and $\mu$ by numerical quadrature. We use OpenGL to compute this value in parallel for all pixels~\cite{engel2006real}. The output 2D x-ray images have a pixel resolution of $256\times 256$. They are in linear physical units, \ie no gamma curve is applied.
Orthographic projection along a random view direction $\mathbf d$ from the positive $z$-hemisphere is used to generate the per-pixel rays. The restriction to the positive $z$ hemisphere is chosen to resolve the following ambiguity in our image formation model: X-ray images taken from direction $\mathbf d$ are identical to x-ray images taken from direction $-\mathbf d$, and consequently two 3D density volumes, where one is flipped along $z$ in camera space have the same image.
Additionally, for each view direction the corresponding images were mirrored in a vertical direction, such that we obtain two instead of one images. Thus, overall $1,332\times 2=2,664$ x-ray images are generated for each species adding up to a total of $2,664\times 175=466,200$ x-ray images, and thus also samples in our dataset, to be made publicly available upon publication.

\mysection{Single-image Tomography}{OurApproach}
To address the single-image tomography problem, we have designed a CNN architecture (\refFig{Architecture}) to learn the mapping from a 2D x-ray image (\refFig{Architecture}, left) to a 3D volume (\refFig{Architecture}, right) from many examples of 3D volumes that are paired with 2D x-rays (\refFig{Samples}).

At \change{deployment time}, the input is a 2D x-ray of arbitrary spatial pixel resolution, \ie higher than the 2D images in the training set, and output is a 3D density volume with the same spatial resolution and a depth of $128$ slices. This is achieved in two stages: a \emph{network} (\refSec{Network}) and a \emph{fusion} (\refSec{Fusion}) step. Input to the network step is the 2D x-ray image re-sampled to $256\times 256$. Output is a $128\times 128 \times 128$ density volume. The fusion step (yellow block in the very right of \refFig{Architecture}) combines this coarse 3D representation with the full-resolution 2D x-ray image (dotted line) into the final resolution. This step is simple enough to be done on-the-fly without the need to even hold the full result in (GPU) memory.

\mysubsection{Network}{Network}{
The network step uses a deep CNN \cite{lecun2015deep}.
The overall structure is an encoder-decoder with skip connections~\cite{ronneberger2015u,long2015fully} and residual learning~\cite{he2016deep}. 
An overview  of our network can be seen in \refFig{Architecture}.
We will now detail some of its design aspects.

\paragraph*{Encoder-decoder.}
The purpose of an encoder-decoder design is to combine abstraction of an image into an internal representation that represents the information contained in the training data (encoder) with a second step (decoder) that applies this knowledge to the specific instance.

To combine global and local information, the network operates on different resolutions~\cite{long2015fully}: the first part (orange blocks in the left half of \refFig{Architecture}) reduces spatial resolution and produces more complex features, as seen from the decreasing horizontal block size and increasing vertical block size in \refFig{Architecture}. 

The right half increases resolution again (blue blocks), but without reducing the feature channel count, as is typically done when the output only has a low number of channels. Spatial resolution is increased by a deconvolution (or up-sampling) unit~\cite{cciccek20163d}.
This deconvolution combines the information about the existence of global features with spatial details in increased resolution.

We found the symmetric encoder-decoder to work best when combined with additional steps before (left grey blocks) and after changing resolution (right pink block). A minimal resolution of 8 provides the best trade-off: larger or smaller minimal sizes result in a larger error in our experiments.

\paragraph*{Skip connections.}
To share spatial details of some resolution at some level on the convolutional part with the same resolution on the de-convolutional part we make use of skip-connections~\cite{long2015fully} (also called cross-links, shown as bridging arrows). These convert fine details in the input 2D image into details of the output 3D volume.
Skip connections allow to use high-resolutional spatial layout to locate features, such as on (3D) edges.

\paragraph*{Residual.}
Furthermore, we use residual blocks to increase the learnability~\cite{he2016deep}. Instead of learning the convolutions directly, we only learn the additive residual and add in the identity. This is seen in the definition of the 3-residual block (dark gray) in \refFig{Architecture}: it combines 3 basic blocks (light gray) with a residual link that provides a ``detour'' resulting in the identity mapping. This does not change the networks expressiveness, but significantly helps the training.

\paragraph*{Convolution.}
The CNN learns image filters of compact support that make the approach scalable to input images and volumes with a sufficient resolution. Convolution (pink block) is typically accompanied by a batch normalization and a ReLU non-linearity. All three steps form a basic block (light gray). 
Usually neural networks map from 2D to 2D or 3D to 3D. However, in our case it is different as 2D to 3D is required. Trivially, one \change{may} think to encode the depth dimension of the volume as the third dimension. This appears attractive as the result is fully convolutional: features deeper in the image/volume are computed in the same way as if they were shallow. However, since the input is 2D and the task is to find the 3D mapping this is not applicable. Therefore, in our case the third volume dimension is encoded as individual feature channels. Consequently, the design increases the number of feature channels from 1 to 256 and retains this number until the end where it is decreased to the output resolution of 128 as seen right in \refFig{Architecture}.
In other words: a network fully convolutional along $z$ would produce the same result for every $z$ slice as nothing ever changes, which is clearly not desirable. 
Future work however could explore switching from feature channels along $z$ to convolutions along $z$ in later steps of a network.

\paragraph*{Learning.}
As our network aims to solve a regression task the loss calculation comprises of a simple $\mathcal L_2$-norm (Euclidean Loss) between the 3D voxels. We train our network using Caffe~\cite{jia2014caffe} in version rc5, and exploit four Nvidia~Tesla~K80 accelerator cards which brings training time down to roughly one day.

\mysubsection{Fusion}{Fusion}
Fusion combines the coarse-resolution 3D result of the previous step into a 3D volume with full spatial resolution. This is based on the intuition, that the overall 3D structure is best captured by the ``intelligence'' of a neural network, while the fine details are more readily available from the high-resolution 2D x-ray image.

Fusion proceeds independently for every pixel in the high-resolution image as follows. Recalling the definition of $\alpha$ from \refEq{Transparency}, we note, that while the loss encourages the inferred densities, say $\bar\mu_i$ of slice $i$ to be close to the ground truth densities $\mu_i$, nothing forces their composition $\bar\alpha$ to be close to the input $\alpha$. This is not surprising, as we \change{do not} know the ground-truth values $\mu$ at test time.
However, we know that they have to combine to $\alpha$ and that the value $\Delta_\alpha=\bar\alpha-\alpha$ is the \emph{transparency error} of our reconstruction. Based on the Beer-Lambert equation, we can compute the \emph{density error} of this as $\Delta=\log(1-\Delta_\alpha)$.
The idea of fusion is to distribute this density error to arrive at new density values $\hat\mu_i$, such that these compose into the correct value $\alpha$ again.

While we would need to know the ground truth to do this correctly, many policies to distribute the error are possible. Consider --- for illustrative purpose --- blaming the entire error on the first slice $\hat\mu_1=\bar\mu_1-\Delta$. It is worth noting how this would result in the correct 2D x-ray, but from a novel view it would show an undesirable ``wall'' of density in front of the object. Instead, one could distribute the error evenly across all $n$ slices, as in $\hat\mu_i=\bar\mu_i-\Delta/n$. This, as any other convex combination of the error, will produce a correct x-ray which will already be much more usable than the first policy. Regrettably, it will also create density in areas that the network has correctly identified as empty, such as the void around each object. This observation leads to the intuition behind the policy we finally suggest, that should change density proportional to density. This is achieved, by setting
\begin{equation}
\hat\mu_i=\bar\mu_i-\Delta
\frac
{\bar\mu_i^\beta}
{\sum_1^n\bar\mu_i^\beta},
\end{equation}
where $\beta=2$ is a sharpness parameter to weight denser areas more.

\mysection{Evaluation}{Evaluation}
We have evaluated the proposed network architecture both on the validation subset of our synthetic dataset, where a ground truth is available, as well as on real images where we do not have access to the ground truth. Furthermore, our approach is compared to three baseline alternatives (\refSec{Competitors}) using two metrics (\refSec{Metrics}).

\mysubsection{Alternative Approaches}{Competitors}
We compare the proposed network architecture (\textsc{Our}) to three alternative approaches which are capable of deriving a 3D volume from a 2D x-ray image.
We refer to those approaches as the nearest-neighbor (\textsc{NN}), the oracle approach (\textsc{Oracle}), and the method of Wenger~\etal\shortcite{wenger2013fast} (\textsc{Wenger}).
We describe them briefly in the following paragraphs.

\mycfigure{ComparisonCombined}{Comparing different methods \emph{(columns)} on different x-rays \emph{(rows)}.
Different methods in different columns are coded as colors also used in the quantitative results found to the right, where we show the numerical results according to SSIM and $\mathcal L_2$ (less is better).
We see that ours is similar to a reference while a real competitor for single-image tomography cannot achieve this.
An oracle or NN method produces plausible skulls, but not the skull in the input x-ray.
This manifests as larger error according to both metrics.
}

\paragraph*{Nearest neighbor.} The nearest neighbor approach uses the input 2D x-ray image to find the most $\mathcal L_2$-similar 2D x-ray in the training data and returns the 3D volume belonging to this same sample. While such a method is feasible in theory, it is very far from practical to remember all 3D volumes and all 2D x-rays as the storage requirement is in the range of \change{terabyte}. Furthermore, the search time would be in the order of a couple of days, whereas our approach only requires less than a second to execute. Nevertheless,  outperforming such a method shows that the problem cannot be solved by memorizing the training data, even if it was feasible.

\paragraph*{Oracle.} The oracle approach simply returns the 3D volume from the training dataset that is closest to the ground truth solution. Note that this ignores the x-ray completely. This is a completely hypothetical method, as it requires to know the ground truth 3D volume of the input, which is not available in practice. Nevertheless, outperforming such a method shows an upper bound on what any memorizing encoding could ever achieve, as no memorization of the data can be better than the data itself.

\paragraph*{Wenger \etal.}
Wenger~\etal have demonstrated single-image tomography for the case of planetary nebulae \cite{wenger2013fast}.
While it clearly has different assumptions and a different objective where it is producing convincing results, it is still the method closest to our objective that we are aware of.
They phrase the problem as an optimization with special constraints such as sparsity and symmetry.
\change{Their method assumes the medium to be emission-only while ours is absorption only}.
Their original implementation was run on our x-ray images.
Due to computation time, we could not run \textsc{Wenger} on the full validation set and will limit ourselves to a representative choice of three images.

\mysubsection{Metrics}{Metrics}
In order to facilitate the comparison, we use two metrics: one 3D volume metric, and one 2D image metric.

\paragraph*{3D Volume Metric.}
For the volume metric $\mathcal L_2$ as used in training \change{is} employed.
The volume metric can account for errors, independent of view point, lighting, iso-value or any other rendering parameters, but is often not well-correlated with the perceived quality of a reconstruction that is dominated for ``what's in'' for the final image.
Smaller $\mathcal L_2$ values of course are better.

\paragraph*{2D Image Metric.}
For computing the image metric, the two volumes are rendered using a canonical setting and the resulting images compared. Rendering is done from a camera identical to the x-ray view (orthographic), but in a different modality: we use iso-surface ray-casting, image based lighting with ambient occlusion and slight specular shading. This is typical for volume visualization and used to visualize our results as well. The resulting images are then compared using DSSIM~\cite{wang2004image}, where again smaller values are better.

\mysubsection{Synthetic Data Evaluation}{Synthetic}
First, we evaluate all approaches using all metrics on synthetic x-ray images, where the ground truth 3D volume is known. For our validation dataset, we find that our approach consistently and significantly performs better than all others according to both metrics.

\mycfigure{ErrorPlot}
{
Error means \emph{(a,c)} and distributions \emph{(b,d)} across our test set for three methods (colors) and two metrics ($\mathcal L_2$ left, SSIM right).
}

\paragraph*{Quantitative results.}
Our mean $\mathcal L_2$ error of $.051$ is significantly ($p<.0001$, paired $t$-test) better (smaller) than the \textsc{NN} method with a mean of $.067$ and the \textsc{Oracle} method $.060$.
The mean and confidence intervals are seen in \refFig{ErrorPlot}, a.
When plotting the distribution of errors in \refFig{ErrorPlot}, b, we further see that no method fairs better than our approach in any regime.
The picture is stronger in DSSIM, where our mean error of $.097$ is significantly better (smaller) than both \textsc{NN} and \textsc{Oracle} methods with means of $.109$, resp.\ $.110$ (both $p<.0001$, in a paired $t$-test). 
We have added \textsc{Wenger} to all plots, despite having only three samples as a rough indication of performance. While it is originally designed for a different purpose, it is the closest competitor we are aware of.
We see that the error is slightly larger than baseline methods using our data.

\paragraph*{Qualitative results}
Finally, the quality is best inspected by comparing our results in re-rendering, cut-away or stereo applications to the ground truth as seen in \refFig{ComparisonCombined} that shows all approaches compared.
We see that \textsc{Oracle} and \textsc{NN} produce volumes that look plausible, but do not really match the input image.
This can be seen from the error bars to the right which are for each individual sample (row) following the color coding of the methods (columns).
Finally, \refFig{SyntheticResults} show more results of our approach, including novel views.
The supplemental materials show the full validation dataset following the protocol of \refFig{SyntheticResults}.

\paragraph*{User study}
When showing the 10 best results according to the SSIM metric (\refFig{SyntheticResults} rows continue) of either GT and ours to $N=27$ na\"ive \change{subjects)} using iso-surface renderings in a time-randomized two-alternative forced choice (2AFC) protocol and asking ``if the image is real", the correct answer was given in $46.3\,\%$ of the cases ($p<.023$, binomial test).
While this might indicate \change{subjects} did not understand the task it could also mean that there is at least no obvious criterion to separate our results from GT.

\mycfigure{Resolutions}
{
Visual comparison of down- and up-sampled volumes (8, 16, 32, 64) to the original volume resolution of 128.
}

\paragraph*{Effect of Slice Count.}
In order to see, if the network is able to reconstruct 128 slices properly, rather than simply interpolating between the slices the 3D output was firstly down-sampled along the depth dimension and instantly up-sampled again to 128 in order to simulate interpolated volumes for the dimensions 8, 16, 32 and 64 respectively. Then each volume was rendered and compared to the original output volume of depth dimension 128 as seen in \refFig{Resolutions}. There are big differences for resolution 8 and 16. For the resolutions 32 and 64 the differences are not that strong in numbers anymore. However, as seen in \refFig{AnglesLowResolution},a there are still differences which means the task performed by the network exceeds interpolation.

\myfigure{AnglesLowResolution}
{
\emph{a)} DSSIM error of down- and up-sampled volumes (8, 16, 32, 64) compared to the original volume resolution of 128.
\emph{b)} DSSIM error across our test set for four different angles: top, front, side, other.
For \emph{(a)} and \emph{(b)}, less is better.
}

\paragraph*{Different views.}
As the 3D volumes are globally aligned before we choose a random view we can analyze the effect of view direction on the error (\refFig{AnglesLowResolution}, b). There are no big differences between different views regarding the output quality. However, x-rays from the front seem to be easier for the network than from other angles, whereas x-rays from the side yield the worst results.

\myfigure{Fusion}{Comparing fusion \emph{(Top)} and no fusion \emph{(Bottom)}.
The re-synthesized x-ray (shown) is fully identical to the input x-ray (not shown) while the iso-surface looks more detailed and remains plausible.}

\paragraph*{Effect of Fusion}
A comparison of the obtained results with and without using the described fusion strategy is shown in \refFig{Fusion}. We note that fusion does not only ensure that our result produces a density-decomposition of the input that is seamlessly compositing into the input again, but also allows the arbitrary handling of high spatial resolutions that would be infeasible to tackle in practice for current CNNs due to the massive amount of data. We would also like to point out that the fusion will never produce more than the $128$  as no additional information in the depth dimension is available from the input 2D x-ray's transparency $\alpha$.

\mysubsection{Real-World Data Evaluation}{Real}
Next, we have applied our network to real-world x-ray images, we have obtained from on-line repositories. Here, we do not have the corresponding 3D volume so quantitative evaluation or rendering from a novel view is not possible.
However, the visual quality is apparent from \refFig{RealResults}.
We see that our approach can extract meaningful three-dimensional structures for unobserved species and real-world x-rays, despite being trained on synthetic images.
The fact that typical x-rays come with gamma compression -- that can only be undone partially -- adds to the difficulty of this task.
Another experiment using real 2D x-rays in combination with real 3D volumes is presented in a separate section \refSec{RealityCheck}.

\myfigure{RealResults}
{
Our results \emph{(bottom)} from real x-ray images \emph{(top)}.
While no reference is available here, the overall shape appears plausible.
}

\mysubsection{Applications}{Applications}
Our approach allows for a couple of interesting computer graphics applications of legacy x-ray images: novel views, stereo, re-rendering and a combination of these.
The supplemental material provides a web-application to explore all combinations 
for 100 samples of the validation dataset.
In the paper, we will constrain ourselves to visualize results only using iso-surface ray-casting with image-based lighting and ambient occlusion with an iso-value of $.1$.

The prime computer graphics application example enabled through our approach is novel view-generation. To this end, the 3D volume is simply input to the image synthesis procedure again, but from a novel point of view. Examples are seen in \refFig{SyntheticResults}.
%
Our approach allows manipulating the obtained 3D density volume, such as cutting away parts (\refFig{Cutaway}). We see that, compared to the ground truth both interior and exterior are predicted.

\myfigure{Cutaway}{Cut-away visualizations of the ground truth \emph{(top)} and our result \emph{(bottom)}.
Note that we reproduce both the surface and the interior of the skull or the back of the jaw.
Cut-aways of the entire validation set are found in the supplemental materials.}
%
The ability to take novel views also allows to produce the two views required for a stereo image as seen in \refFig{Stereo}.

\myfigure{Stereo}{Stereo from x-rays. Anaglyph and wiggly stereo visualization of the entire validation set are found in the supplemental.}

\mysubsection{Reality check}{RealityCheck}
A methodological dilemma is that ultimately we want to know performance on true x-ray images, but regrettably, we do not see a viable way to attain the same amount of real-world x-ray images as can be acquired for synthetic ones. In the following paragraphs we address this challenge with one observation and an additional experiment.

\myfigure{Puzzle}{Which x-rays are real and which are synthetic?}

\paragraph*{Realism of synthetic x-rays.}
First, we note that our synthetic x-ray images are likely similar to the real x-ray images. This is hard to quantify, but a reader is encouraged to try to detect which x-rays in \refFig{Puzzle}, that show both our x-rays and x-rays from the internet are real and which one are synthetic.
When time-sequentially showing 10 space-randomized pairs of real and fake x-ray images to na\"ive subjects in a 2AFC task and asking ``which image is a real x-ray'', the correct answer was given in $50.21\%$ of the cases \ie chance-level at 50\,\% ($N=46$, $p=.037$, binomial test). 
The supplemental materials show these stimuli.
While this is no formal proof of our performance on real-world x-rays, it indicates that at least the differences in x-rays are not easily detected, and that they could be close. This is likely because x-ray transport is less complex (less scattering, no reflection, only absorption) than light transport on the size-scales of our geometry.

\paragraph*{Learning from real-world x-rays.}
Second, we have repeated the entire learning for a restricted subset of x-ray images for which we have explicitly reconstructed both image and volume. In this set, we used our own micro-CT scanner to acquire x-rays of 15 mice (\textit{Mus musculus}), which were then reconstructed into volumes using classical tomography. We split this set into 12 training and 3 test exemplars, produced $2,720$ overall samples and repeated the entire learning procedure explained above.

Qualitative results of this experiment are shown in \refFig{Mice}. Quantitatively, we find, that, despite the restricted setting again we outperform a NN and Oracle comparison (\refFig{ErrorPlotMice}). Again, both our mean and error distributions are better than any competitor for any metric. While it does not show generalization across species, it shows that if the scanning effort is made, our method is applicable to real volumina and x-rays. If we had access to the massive original 2D x-ray image data from UTCT~\cite{UTCT}, a similar experiment could be repeated on a full scale, providing an actual proof. Regrettably, the x-ray data for those scans is not available anymore.

\mycfigure{Mice}{
Applying our approach to recover the 3D internal structure of mice X-rays.
The left image shows the x-ray image.
The middle column show the reconstructed 3D volume rendered using a transfer function.
This can be compared to the reference, shown on the right.
We see that \protect\change{both} the external skin structures shown in blue and the bones shown in orange are present.
} 

Finally, we re-synthesized x-rays from the density volumes, closing the loop in \refFig{ReSynthesis} to compare them to the real x-rays. We find, that resulting images are very close and training with these re-synthesized images leads to similar results as seen in \refFig{ReSynthesis}. This indicates, at least on a small scale, that our approach  has learned the inversion of real x-rays by training on synthetic x-rays.

\myfigure{ReSynthesis}{Re-synthesis of x-rays (\emph{bottom}) from CT scans where the original x-rays are available (\emph{top}).
Both are similar, and it would not be obvious which one is synthesized and which is real.
}

\mycfigure{ErrorPlotMice}{Evaluation on real x-ray imagery of mice, using the protocol of \protect\refFig{ErrorPlot}, allowing for the same conclusion.}

\mysection{Conclusion}{Conclusion}
We have demonstrated the first application of deep learning to reconstruct 3D volumes from single 2D x-ray images. After suggesting a novel dataset for evaluation and testing, we have devised a deep CNN that can produce full 3D volumes. We suggest a specialized fusion step, that allows training on low resolution examples, yet transferring the outcome to high-resolution input. A similar approach could be applicable to other conditions that are limited by the sheer amount of data (video, 3D video, light field (video), etc.). Our method was tested, both on synthetic and real images, allowing for novel applications such as free viewpoint, viewing of legacy x-ray footage, stereo x-ray imagery or re-rendering in new modalities.

We have only looked at one specific instance of learning volumes from images. Our approach was learned on and tested with skulls, which form a prominent and intriguing class, but are by far not the only class. Our experiments on CTs of real mice indicate that the method can be trained on both synthetic and real data. Other geometry worth reconstructing could be clouds or smoke in applications such as weather forecast, or x-ray images obtained from security scanners at airports. We have chosen an absorption-only transport model that suits x-rays. For photos, \eg of clouds or smoke, an emission or emission-absorption model would need to be learned. We imagine our setup would trivially extend to this case. As x-rays are known to be dominated by single-scattering, future work would need to account for multiple scattering in other modalities.

Finally, 3D-volumes-from-X in other modalities such as PET or ultrasound, will be subject to phenomena not typically modeled in graphics, \eg diffraction, requiring even more refined synthesis of training data providing excellent avenues of future research \change{eventually producing a generalized image-to-image translation \cite{isola2016image}, mapping from 2D to 3D images}.

\mycfigure{SyntheticResults}{
Results of different approaches (\emph{columns}) on different species \emph{(rows)} in our synthetic validation data.
The first column shows the input 2D x-ray image.
The second and third column show a rendering of our result resp.\ the ground truth rendered from the original view.
The last two columns use a novel view.
We see that our approach can recover non-trivial details of the mamalian morphology such as the cheekbones.
The overall shape and surface orientation is plausible as seen from the colors in the shading. Even from novel views our results look convincing, most notably when reproducing holes and cavities not present in any height field.
}

\begin{small}
\vspace{-.2cm}
\paragraph*{Acknowledgements}
We thank
T.\ Leimk\"uhler, O.\ Nalbach, M.\ Firman and I.\ Oakes for proofreading,
U Texas,
the U T\"ubingen HPC group,
the state of Baden-W\"urttemberg for bwHPC, DFG grant INST 37/935-1 FUGG,
LRS for a new hypothesis,
and all mice.
\end{small}

\bibliographystyle{eg-alpha}
\bibliography{bibliography}

\end{document}